\documentclass[journal=jctcce,manuscript=article, layout=twocolumn]{achemso}

\usepackage[utf8]{inputenc}
\usepackage{amssymb}
\usepackage{graphicx}
\usepackage{caption}
\usepackage{subcaption}
\usepackage{float}
\usepackage{amsmath}

\usepackage{xcolor}
\usepackage{hyperref}
\usepackage[ruled,vlined,linesnumbered]{algorithm2e}
\usepackage{xr}

\title{Top-down machine learning of coarse-grained protein force-fields}

\author{Carles Navarro}
\affiliation{Acellera Labs, Doctor Trueta 183, 08005, Barcelona, Spain}

\author{Maciej Majewski}
\affiliation{Acellera Labs, Doctor Trueta 183, 08005, Barcelona, Spain}

\author{Gianni De Fabritiis}
\affiliation{Computational Science Laboratory, Universitat Pompeu Fabra, Barcelona Biomedical Research Park (PRBB), Carrer Dr. Aiguader 88, 08003, Barcelona, Spain}
\alsoaffiliation{Acellera Ltd, Devonshire House 582, HA7 1JS, United Kingdom}
\alsoaffiliation{Institució Catalana de Recerca i Estudis Avançats (ICREA), Passeig Lluis Companys 23, 08010 Barcelona, Spain}
\email{gianni.defabritiis@upf.edu}
\date{}
\graphicspath{ {images/} }

\begin{document}
\maketitle

\begin{abstract}

Developing accurate and efficient coarse-grained representations of proteins is crucial for understanding their folding, function, and interactions over extended timescales. Our methodology involves simulating proteins with molecular dynamics and utilizing the resulting trajectories to train a neural network potential through differentiable trajectory reweighting. Remarkably, this method requires only the native conformation of proteins, eliminating the need for labeled data derived from extensive simulations or memory-intensive end-to-end differentiable simulations. Once trained, the model can be employed to run parallel molecular dynamics simulations and sample folding events for proteins both within and beyond the training distribution, showcasing its extrapolation capabilities. By applying Markov State Models, native-like conformations of the simulated proteins can be predicted from the coarse-grained simulations. Owing to its theoretical transferability and ability to use solely experimental static structures as training data, we anticipate that this approach will prove advantageous for developing new protein force fields and further advancing the study of protein dynamics, folding, and interactions.

\end{abstract}

\section{Introduction}

Molecular dynamics (MD) simulations are a valuable tool for investigating various biomolecular processes, such as protein-protein interactions\cite{interactions}, folding and unfolding\cite{Baxa, Noe, Naganathan}, and protein-ligand binding\cite{ligand}\cite{McGeagh2011, Muller2019, Maffeo2012, Petrov2013}. However, conventional MD methods face limitations in their applicability to these processes, mainly due to high computational costs and large timescales involved. Coarse-grained protein modeling methods have emerged as a potential solution to this challenge\cite{Peter2013, Pantano2019, Voth2013, Gregory2005, Noid2013, CLEMENTI200810, CLEMENTI2000937}. These methods capture the essential dynamics of the system with reduced degrees of freedom, enabling the exploration of longer timescales, but require careful design of the coarse-grained representation and force field for accurate mapping to a lower-dimensional space\cite{Clementi2018, Noid2015, Noid2020}.

Early efforts in the development of coarse-grained (CG) protein potentials laid the foundation for knowledge-based (KB) protein models. Early KB works \cite{Miyazawa1985, Sippl1990} established significant progress in creating statistical potentials from protein data bank (PDB) structures. Subsequent KB studies \cite{Maiorov1992, Mirny1996, Liwo2002, Papoian2004, Sterpone2014} furthered this field by designing CG potentials to stabilize PDB structures. Alternatively, bottom-up approaches for parameterizing a CG model from PDB conformations have also been utilized. These methodologies postulate that PDB conformations dominate the equilibrium ensemble so can be used to determine transferable interaction potentials for CG protein models, with statistical physics approaches that treat many-mody structural correlation \cite{Mullinax2010}, maximizing the likelihood that the CG model samples PDB conformations, as demonstrated in the Bayesian approach\cite{Chen2017} or with the relative entropy approach \cite{Shell2008}.

In recent years, the adoption of machine learning algorithms in molecular dynamics has gained traction, driven by the increasing availability of experimental data, computational power, and auto-differentiation software\cite{Gkeka2020, Zhang2018, Wang2019, WWang2019}. A particularly promising application of machine learning in MD is the development of neural network potentials (NNPs)\cite{Behler2007}. NNPs can effectively model many-body interactions and represent potential energy surfaces\cite{Tkatchenko2020}, and have been employed to construct machine-learned coarse-grained force fields for biomolecules\cite{ani, Brooke, Wang2019, Stefan2021, Maciej2023, torchmdnet}. 

The development of CG machine-learned NNPs for proteins generally adopts a bottom-up approach, which attempts to reproduce the reference fine-grained statistics. Initial studies focused on directly learning potentials from high-fidelity simulations through variational force-matching\cite{Brooke, Wang2019, Maciej2023}, while more recent research has explored data-efficient strategies, such as flow-matching\cite{Kohler2023} and denoising diffusion probabilistic models\cite{Berg2023}. This bottom-up approach has the advantage of preserving the thermodynamics of the projected degrees of freedom. However, it also necessitates a large quantity of all-atom 3D conformations and their corresponding energies and forces sampled from the equilibrium distribution for training the machine learning model. This requirement can be computationally expensive and may result in poor extrapolation in regions of conformational and sequence space where data is scarce\cite{Noe2020}.

On the other hand, an alternative approach for learning potential energy functions has been demonstrated by Greener et al.\cite{greener} through the application of end-to-end differentiable molecular simulations (DMS). However, this method faces challenges when applied to medium-to-large proteins, as it requires a significant amount of memory to store all simulation operations during the forward pass, which are then used in the backward pass. This can lead to exploding gradients, causing instability during training as the accumulated gradients result in large updates in the neural network weights. To address this issue, the differentiable trajectory reweighting (DiffTre) method has been developed and applied to learn NNPs for atomistic systems in a more memory-efficient manner\cite{Thaler2021}. Bypassing the differentiation of the entire trajectory through the MD simulation for time-independent observables, DiffTre enables the learning of top-down coarse-grained potentials.

In this work, we build upon the advantages of DiffTre and demonstrate its applicability for training NNPs of coarse-grained proteins using experimental structures and short differentiable MD trajectories from various proteins. Our approach allows us to train an NNP using differentiable trajectories and uncorrelated states while circumventing the need to save all simulation operations for the backward pass. As a result, our approach is considerably less memory-intensive while retaining the performance of DMS. We apply the proposed methodology to learn two distinct NNPs: one for 12 fast-folding proteins (FF-NNP), which can be utilized to recover their native conformations, and a general NNP (G-NNP) trained with a much larger dataset of computationally solved structures, which demonstrates the capability of extrapolating and folding proteins outside of the training set.

\section{Methods}
\subsection{Coarse Grained Molecular Dynamics}

The design of a coarse-grained model starts with the definition of the variables that should be preserved after the dimensionality reduction. In this study, to reduce the dimensionality of system $\textit{R} \in \mathbb{R}^{3N}$, we establish a linear mapping $\Xi: \mathbb{R}^{3N} \rightarrow \mathbb{R}^{3n}$, projecting it onto a lower dimensional representation $r \in \mathbb{R}^{3n}$. We transform the all-atom representation into a $C_{\alpha}$ atom representation, with retained atoms referred to as coarse-grained "beads". Each $C_{\alpha}$ bead is assigned a bead type based on amino acid type, resulting in 21 unique bead types, which are identified by distinct integers.  

Molecular dynamics simulations are employed for training and testing the NNP. We utilize TorchMD \cite{Stefan2021}, a fully differentiable MD package written in PyTorch \cite{pytorch}, enabling the execution of multiple parallel trajectories. To confine the space explored by the dynamics and incorporate physical information, we apply a prior energy function $U_{\lambda}(r, \phi)$. TorchMD integrates prior energy terms and the NNP to compute total potential energy. Consequently, the potential energy function is decomposed into a prior potential $U_{\lambda}$ and a neural network potential $U_{\theta}$, with the total potential energy given by:
\begin{equation}
        \label{eq:potential}
      U_{\theta, \lambda}(r, \phi) = U_{\theta}(r) + U_{\lambda}(r, \phi),
\end{equation}

where $U_{\theta}$ represents the potential energy derived from the NNP, parameterized by parameters $\theta$. The network is a graph neural network with a SchNet-based \cite{Schutt2018} architecture, a continuous-filter convolution neural network capable of modeling atomic interactions. The implementation, available in the Torchmd-NET package \cite{torchmdnet}, is defined by Majewski \textit{et al.} \cite{Maciej2023}. The prior energy $U_{\lambda}$ is parameterized by constant parameters $\lambda$, which can be decomposed into three terms: a pairwise bonded term to prevent chain breaking, a nonbonded repulsive term to avoid structure collapse into overlapping beads, and a dihedral term to enforce chirality in the system. The Supporting Information provides a detailed description of the prior energy terms. The total energy of the system is then expressed as:

\begin{equation}
    \label{eq:pot_exp}
\begin{split}
    U_{\theta, \lambda}(r, \phi) = &U^{harmonic}_{\lambda}(r) + U^{repulsive}_{\lambda}(r) + \\ &U^{dihedral}_{\lambda}(\phi) + U_{\theta}^{NNP}(r).
\end{split}
\end{equation}

To compute the forces, TorchMD computes analytically the forces from the priors and obtains the NNP forces with an autograd PyTorch call on the energy term computed with the NNP. 

\subsection{Differentiable Trajectory Reweighting}
We implement a version of DiffTre \cite{Thaler2021} in PyTorch to train the NNP, facilitating parallel and distributed training across multiple GPUs and nodes. The package is modular, allowing training for any molecular system and using any experimental observable as a training objective. DiffTre is used to match \emph{K} outputs of molecular dynamics simulations to experimental observables. In this study, we focus on the folding of coarse-grained proteins, using conformations $\textbf{r}_0 \in \mathbb{R}^{3n}$ of proteins in their native states as experimental observables, where $n$ denotes the number of beads in the system. 

In the context of our work, a state denoted as $S_i$, represents a specific configuration of the system at a given time in our simulations. More specifically, $S_i$ is a multidimensional entity that encapsulates both the spatial coordinates $\textbf{r}{i} \in \mathbb{R}^{3n}$ of the system, and the potential energy $U_{\lambda, \hat{\theta}, i}\in\mathbb{R}$ of the system.

As illustrated in Algorithm \ref{alg:1}, we simulate multiple trajectories of length $N$ in parallel, sampling $K$ uncorrelated states $S_i$ from each trajectory $\{ S_i \}_{i=1}^N$. For each state, we compute the root-mean square deviation (RMSD) between the state's coordinates and the native conformation coordinates. Subsequently, the weighted RMSD ensemble average can be calculated as

\begin{equation}
    \label{eq:ensemble}
    \langle RMSD \rangle =  \sum_{i=1}^K w_i RMSD(\textbf{r}_0, \textbf{r}_i),
\end{equation}
\begin{equation}
    \label{eq:weights}
    w_i = \frac{e^{- \beta (U_{\lambda, \theta} (r_i) - U_{\lambda, \hat{\theta}} (r_i))}}{\sum_{j=1}^K e^{- \beta (U_{\lambda, \theta} (r_j) - U_{\lambda, \hat{\theta}} (r_j))}},
\end{equation}
where $U_{\lambda, \hat{\theta}}$ denotes the potential energy calculated with the reference parameters that generate the trajectory, $U_{\lambda, \theta}$ represents the potential energy calculated with the parameters to be updated, and $\beta = 1/(k_BT)$, with $k_B$ being the Boltzmann constant and T the temperature. Note that before the first backward pass, $\theta = \hat{\theta}$, thus $w_i = 1/K$. 

The algorithm's objective is to minimize a loss function $\mathcal{L}_{\theta}$, which in turn minimizes the ensemble average of the RMSD function. During optimization, the ensemble average RMSD between states sampled from short MD simulations and the native conformation of each protein is minimized. To avoid overfitting on proteins that are easier to optimize for the network, we employ a margin-ranking loss
\begin{equation}
    \mathcal{L}_{\theta} = \frac{1}{M}\sum^M_{k=1}[max(0, \langle RMSD \rangle + m)],
     \label{eq:loss}
\end{equation}
where M represents the batch size and $m$ denotes the margin. For example, when the margin is set to $-1$ \AA, if the RMSD ensemble average is lower than 1 \AA, the loss is set to 0 \AA, and the network parameters will not be updated. Training is considered to have reached convergence when the training loss remains constant within a specified error range, and further optimization is unlikely to yield significant improvements. 

\begin{algorithm}[t]
\SetAlgoLined
\SetAlgoNoLine
\textbf{Input:} Prior energy function $U_{\lambda}$, Neural Network Potential $U_{\theta}$, dataset of native conformations $\mathcal{D}$, Batch size $M$, Trajectory length $N$, Number of states to sample $K$, learning rate $\eta$, margin $m$ \\
$\hat{\theta} \longleftarrow  \theta$ \\
\Repeat{Convergence}{
Draw a minibatch of samples \{ ${r_0^1, \dots, r_0^M}$ \} from $p_{data} \sim \mathcal{D}$. \\
Sample $K$ states for every sample of the minibatch \{$\{S_1, \dots, S_K\}_1, \dots, \{S_1, \dots, S_K\}_{M}$\} from $q_{\lambda, \hat{\theta}} \propto	exp(-U_{\lambda, \hat{\theta}})$ (Using Molecular Dynamics trajectories of length $N$) \\

Update the neural network parameters by stochastic gradient descent:
\begin{flalign*}
&w_i = \frac{e^{- \beta (U_{\lambda, \theta} (r_i) - U_{\lambda, \hat{\theta}} (r_i))}}{\sum_{j=1}^K e^{- \beta (U_{\lambda, \theta} (r_j) - U_{\lambda, \hat{\theta}} (r_j))}}\\
&\langle RMSD \rangle \simeq  \sum_{i=1}^K w_i RMSD(\textbf{r}_0, \textbf{r}_i)\\
&\mathcal{L}_{\theta} = \frac{1}{M}\sum^M_{k=1}[max(0, \langle RMSD \rangle + m)]\\
&\theta \longleftarrow \theta - \eta \nabla_{\theta} \mathcal{L}_{\theta}
\end{flalign*}
$\hat{\theta} \longleftarrow \theta$
}
\caption{Training DiffTre for Protein Structure Potentials}
\label{alg:1}
\end{algorithm}

\subsection{Markov State Models}

Markov State Models (MSM) \cite{Husic2018, Prinz2011, Singhal2004, Noe2008, Pan2008} are employed to analyze the CG simulations and compare them with their corresponding all-atom simulations. MSMs can describe the entire dynamics of a system by partitioning it into $n$ discrete states. For a system to be Markovian, it must be "memoryless", meaning that future states depend only on the current state. In the case of Markovian systems, such as MD simulations, a transition probability matrix can be constructed, characterized by the $n$ states and the lag time $\tau$ at which the system's state is recorded. From this matrix, state populations and conditional pairwise transition probabilities can be obtained, and the state populations can be converted into free energies. 

In this work, we employ time-lagged independent component analysis (TICA) \cite{Perez2013, Schwantes2013} to project the high-dimensional conformational space into an optimally reduced low-dimensional space. Following this, the resulting space is discretized using K-means clustering for MSM construction. We featurize the simulation data into pairwise $C_{\alpha}$ distances and use TICA to project the data onto the first four components. For the coarse-grained (CG) data, we adopt the approach presented by Maciej et al. \cite{Maciej2023}, using the covariance matrices of the all-atom molecular dynamics (MD) to project the first three components. This method ensures consistency with established methodologies and facilitates further analysis. 

We use MSMs in the coarse-grained trajectories because we had them already for the all-atom trajectories and to evaluate the shape of the folding basins.  However, we have no expectation that the thermodynamics or kinetics of these coarse-grained simulations have anything to do with the original ones as the training methods do not preserve these quantities. It can be interpreted as a way to get stable states, which we can take as predictions for the native structure. By comparing the most probable macrostate to the protein's native conformation, we can evaluate the predictive capabilities of our CG model. For clustering the data we apply the Pairwise Constrained Component Analysis \cite{Mignon2012} (PCCA) algorithm.

\subsection{Datasets}
The first dataset comprises the crystal structures of 12 fast-folding proteins, previously studied by Lindorff-Larsen \textit{et al.}\cite{lindorff2011fast} (using all-atom MD) and Majewski \textit{et al} \cite{Maciej2023} (using Machine Learned CG MD). These proteins exhibit a variety of secondary structure elements, including $\alpha$-helices and $\beta$-sheets. 

The second dataset, used for training the general model, was created by searching the AlphaFold Database \cite{AlphaFoldDB}. This dataset contains 14,871 proteins between 50 and 150 residues, with computationally predicted structures solved using AlphaFold2 \cite{Jumper2021}. The selected structures in this dataset are high-quality predictions, with a global predicted local-distance difference test (pLDDT) higher than 90, and are clustered at 50\% sequence identity using Usearch \cite{Usearch}. We also removed any hits with more than 20\% sequence identity to the fast-folding proteins used for testing. This approach ensures a diverse representation of protein structures in the dataset, allowing the NNP to generalize effectively to new and previously unseen protein structures. By incorporating AlphaFold2-predicted structures into our dataset, we significantly increase its size compared to using experimentally solved structures alone. 
\begin{figure*}[t!]
      \centering
      \captionsetup{width=1\linewidth}
      \includegraphics[width=\linewidth]{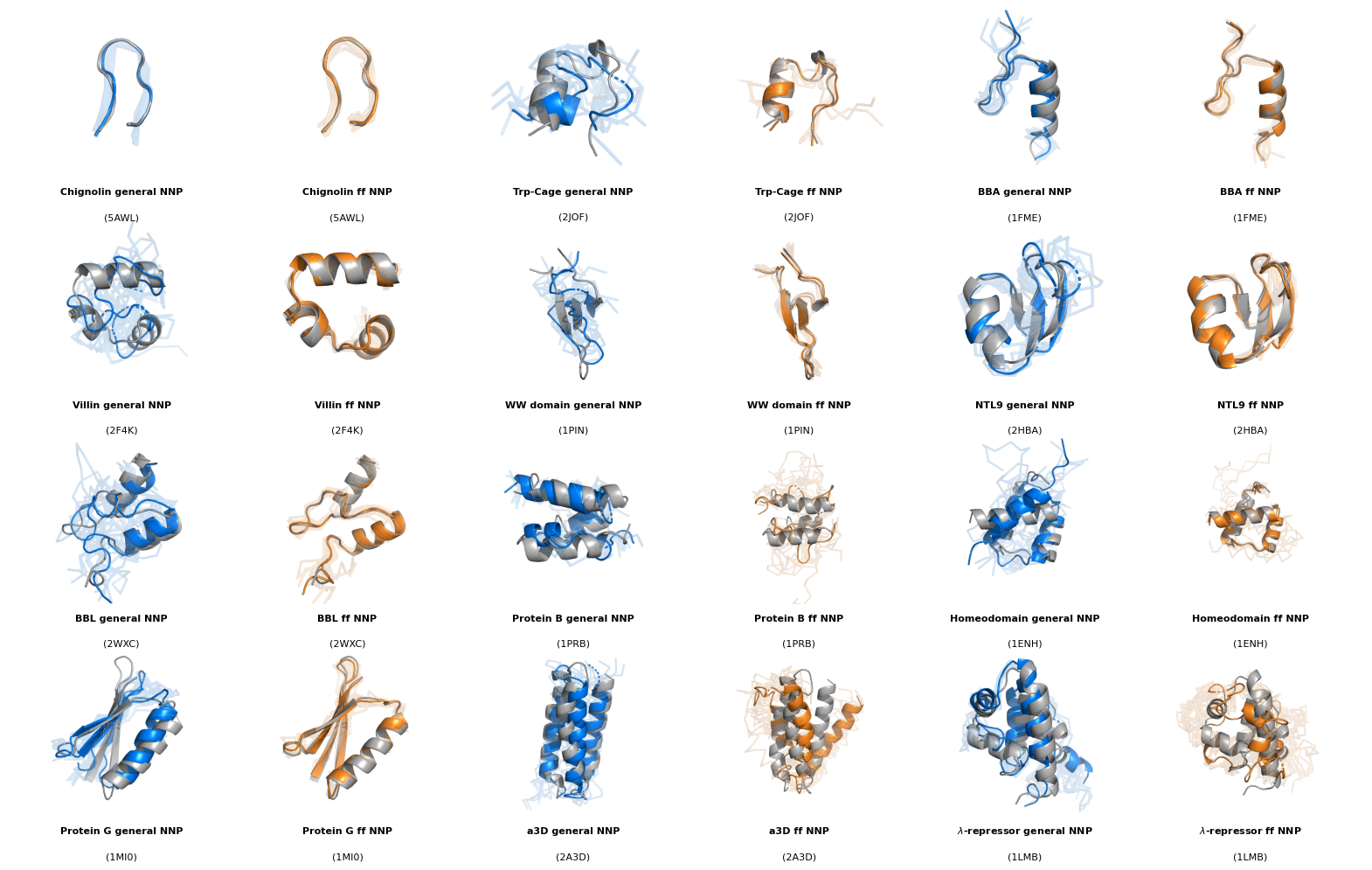}
      \caption{Representative structures sampled from the minimum RMSD macrostate from coarse-
grained simulations of the 12 fast-folding proteins. For each protein we show the experimental structure (gray), the selected conformations from the general NNP (blue) and the selected structures from the fast folders NNP (orange). 10 structures are randomly selected from each model's minimum RMSD macrostate, with the minimum RMSD one highlighted and the other as transparent shadows.}
     \label{fig:conformations}
\end{figure*}
\section{Results}
\subsection{A Neural Network coarse-grained potential learns the structure of fast-folding proteins.}
\label{section:ffNNP}
We trained the FF-NNP using the dataset of 12 fast-folding proteins. For training, the learning rate was set to $\epsilon = 0.0001$, and the loss function was defined by Equation (\ref{eq:loss}). The simulation temperature was set to 298 K, and the timestep was 5 fs. We employed trajectories of 1024 steps, with 128 states used for reweighting each trajectory. The margin was set to $m = -1.0 ; \text{\AA}$, and the mini-batch size was 12. 
Compared to bottom-up approaches \cite{Maciej2023, Kohler2023, Berg2023, Wang2019, Brooke}, our training process did not require generating expensive reference all-atom data beforehand, and the training took only 5 hours on a single NVIDIA GeForce RTX 2080 GPU. Despite this, the model successfully folded most of the proteins and stabilized their native conformations. 

To validate the fast folders' NNP, we performed coarse-grained molecular dynamics simulations using the same proteins used for training. In order to ensure a more comprehensive exploration of the conformational space, we took advantage of the parallel processing capabilities of TorchMD to initiate multiple parallel trajectories for each protein. Rather than starting all trajectories from unfolded conformations, which might limit the exploration, we diversified our starting points. We used 32 different conformations as starting points, representing a wide array of distinct points in the conformational space. 

These starting conformations were selected to create a wider initial condition set, promoting a broader exploration of the potential energy landscape. While in a typical experimental setup, such a wide range of initial conformations might not be readily available, our computational approach enabled this extended exploration, which we believe is key to a more robust validation of the NNP. Our simulations were conducted with a timestep of 1 fs and a temperature of 298 K, running for a total aggregated time of 64 ns for each protein.

From the MSM analysis, we selected the minimum average RMSD macrostate for each protein, considering it as the native macrostate of the simulation. Table \ref{tab:minMacro} presents the equilibrium probability of this macrostate, along with its mean and minimum RMSDs. The results suggest that the fast folders' NNP simulations successfully recovered the experimental structure for all simulated proteins, except $\lambda$-repressor, a3d, and Protein B (Tab. \ref{tab:minMacro}). Furthermore, the equilibrium probability of all these native macrostates is high, indicating extensive sampling of the native conformation. Representative conformations from the macrostate are shown in Figure \ref{fig:conformations}. For Protein B and $\lambda$-repressor, the lowest RMSD macrostates exhibit high flexibility and do not form any secondary structure. In contrast, for a3d, the secondary structure is recovered, although the tertiary structure is not correctly aligned. 

\begin{table*}[t]
  \centering
  \resizebox{\linewidth}{!} {
  \caption{Minimum average RMSD macrostate statistic from MSM built with CG simulations of the fast-folders, general NNP and all-atom MD. The data shows the average and minimum RMSD of the macrostate, as well as its equilibrium probabilities in percentage (Macro prob.) with standard deviation. In bold the proteins with Mean RMSD $<$ 3.0 \AA.}
  \begin{tabular}{cccccccccccc}
    & \multicolumn{3}{c}{FF-NNP} & & \multicolumn{3}{c}{G-NNP} & & \multicolumn{3}{c}{All-atom} \\
    \cline{2-4} \cline{6-8} \cline{10-12}
    Protein & Min RMSD & Mean RMSD & Macro prob. & &
    Min RMSD & Mean RMSD & Macro prob. & &
    Min RMSD & Mean RMSD & Macro prob. \\
    & (\AA) & (\AA) & (\%) & & 
    (\AA) & (\AA) & (\%) & & 
    (\AA) & (\AA) & (\%) \\
    \hline
    Chignolin & \textbf{0.3} & \textbf{0.9 $\pm$ 0.7} & \textbf{42.5 $\pm$ 0.2} & & \textbf{0.3} & \textbf{1.8 $\pm$ 0.4} & \textbf{24.5 $\pm$ 0.1} & &
    \textbf{0.1} & \textbf{1.0 $\pm$ 0.4} & \textbf{57.5 $\pm$ 0.2}\\
    Trp-cage & 1.1 & 3.2 $\pm$ 1.4 & 5.5 $\pm$ 0.1 & &
    1,9 & 5.2 $\pm$ 1.3 & 1.7 $\pm$ 0.1 & &
    \textbf{0.4} & \textbf{2.5 $\pm$ 0.8} & \textbf{30.1 $\pm$ 0.2}\\
    BBA & \textbf{0.4} & \textbf{1.3 $\pm$ 0.7} & \textbf{30.0 $\pm$ 0.1} & & 
    \textbf{1.7} & \textbf{3.0 $\pm$ 0.5} & \textbf{17.3 $\pm$ 0.1} & &
    1.1 & 3.9 $\pm$ 1.3 & 5.2 $\pm$ 0.2\\
    WW-domain & \textbf{0.4} & \textbf{1.0 $\pm$ 0.6} & \textbf{1.8 $\pm$ 0.2} & &
    2.5 & 6.5 $\pm$ 1.1 & 16.5 $\pm$ 0.1 & &
    \textbf{0.7} & \textbf{2.7 $\pm$ 1.1} & \textbf{45.5 $\pm$ 0.0}\\
    Villin & \textbf{0.4} & \textbf{1.0 $\pm$ 0.5} & \textbf{18.6 $\pm$ 0.1} & & 
    4.3 & 7.1 $\pm$ 0.8 & 30.6 $\pm$ 0.1 & &
    0.5 & 3.4 $\pm$ 1.8 & 69.4 $\pm$ 0.2\\
    NTL9 & \textbf{0.5} & \textbf{0.9 $\pm$ 0.3} & \textbf{16.0 $\pm$ 0.1} & &
    1.6 & 4.2 $\pm$ 0.8 & 11.2 $\pm$ 0.1 & &
    \textbf{0.3} & \textbf{1.6 $\pm$ 0.9} & \textbf{15.3 $\pm$ 0.2}\\
    BBL & \textbf{0.5} & \textbf{1.8 $\pm$ 0.6} & \textbf{4.6 $\pm$ 0.2} & &
    2.5 & 6.3 $\pm$ 1.3 & 36.7 $\pm$ 0.1 & &
    1.2 & 3.1 $\pm$ 0.8 & 5.2 $\pm$ 0.1\\
    Protein B & 4.0 & 8.6 $\pm$ 2.5 & 9.9 $\pm$ 0.1 & & 
    3.7 & 4.5 $\pm$ 0.5 & 11.5 $\pm$ 0.2 & &
    1.2 & 4.4 $\pm$ 1.4 & 30.1 $\pm$ 0.1\\
    Homeodomain & 0.5 & 3.6 $\pm$ 3.9 & 35.0 $\pm$ 0.1 & &
    1.6 & 6.2 $\pm$ 2.1 & 39.2 $\pm$ 0.1 & &
    \textbf{0.3} & \textbf{2.3 $\pm$ 1.5} & \textbf{53.5 $\pm$ 0.2}\\
    Protein G & \textbf{0.5} & \textbf{1.0 $\pm$ 0.4} & \textbf{1.5 $\pm$ 0.1} & & 
    1.3 & 3.6 $\pm$ 1.0 & 12.6 $\pm$ 0.1 & &
    \textbf{0.6} & \textbf{2.9 $\pm$ 1.9} & \textbf{17.1 $\pm$ 0.1}\\
    a3d & 3.4 & 8.5 $\pm$ 2.3 & 5.8 $\pm$ 0.1 & & 
    2.0 & 3.7 $\pm$ 1.0 & 29.8 $\pm$ 0.1 & &
    1.8 & 3.5 $\pm$ 0.7 & 67.9 $\pm$ 0.3\\
    $\lambda$-repressor & 4.9 & 6.8 $\pm$ 1.5 & 0.4 $\pm$ 0.2 & & 
    3.6 & 5.3 $\pm$ 0.6 & 2.0 $\pm$ 0.2 & &
    0.8 & 4.5 $\pm$ 1.2 & 21.9 $\pm$ 0.3 \\
    \hline
    \label{tab:minMacro}
  \end{tabular}
}
\end{table*}

\subsection{A Neural Network Potential trained on a Large Protein Dataset learns beyond training data distribution}

In the previous section, we demonstrated that a folding NNP can be learned for a small set of proteins. In this section, we present the results of the G-NNP, which is trained on a much larger dataset of around 15 thousand protein structures. This enables us to test the generalization capabilities of our approach.  We used a random 90/10\% training/validation split. Furthermore, we built the training dataset such that it did not contain sequences with a sequence similarity greater than 20\% to the fast folders. Thus, we use the 12 fast-folding proteins, providing a well-known but independent test set of our model's ability to generalize to proteins that fall outside of the training set, allowing us to directly compare the results with the FF-NNP and reference all-atom simulations.  

We trained the G-NNP using a learning rate $\epsilon = 0.0005$. Our training utilized trajectories consisting of 100 steps, with each trajectory being reweighted using 20 state and the mini-batch size was 32. With the abundance of data from our large dataset, we were able to set the margin (m) to 0 Å, maintaining an optimal balance and avoiding the issue of overfitting. This parameter configuration was found to yield optimal results when training on large datasets.

Similar to the FF-NNP, we initialized multiple parallel trajectories for each protein. We used the same conditions and starting points as those used in Section \ref{section:ffNNP} with the same total aggregated time. 

\begin{figure*}[t!]
      \centering
      \captionsetup{width=1\linewidth}
      \includegraphics[width=1\linewidth]{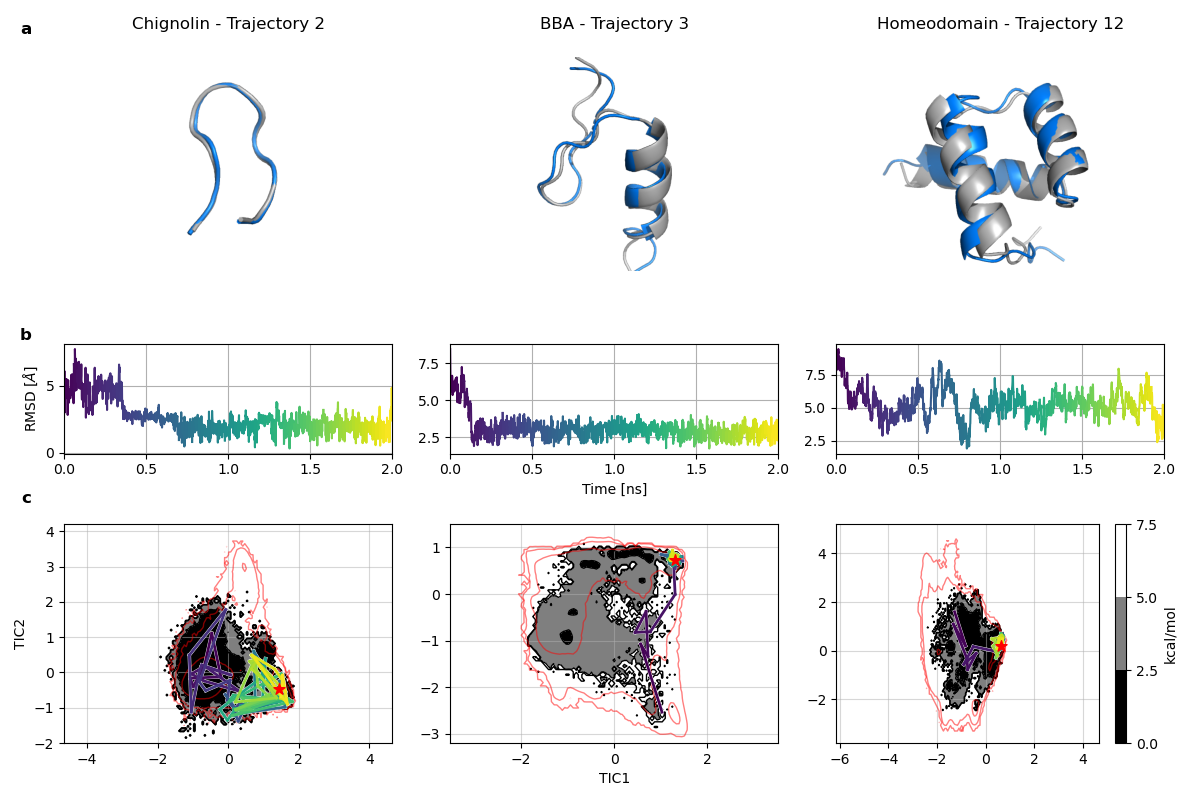}
      \caption{Three individual CG trajectories selected from MD of Chignolin, BBA and Homeodomain. Each trajectory is visualized using different colors ranging from purple to yellow. Each simulation started from an extended conformation and sampled the native structure.  Top panels: Minimum RMSD conformation of the trajectory (blue) aligned with the experimental structure (gray) for Chignolin, BBA and Homeodomain. Middle panels: C$_\alpha$ RMSD of the trajectory with reference to the experimental structure for Chignolin, BBA and Homeodomain.  Bottom panels:  100 states sampled uniformly from the trajectory plotted over CG free energy surface, projected over the first two time-lagged independent components (TICs) for Chignolin, BBA and Homeodomain. The red line indicates the all-atom equilibrium density by showing the energy level above the free energy minimum with the value of 7.5 kcal/mol. }
     \label{fig:folding_general}
\end{figure*}

Results from the MSM analysis reveal that for specific proteins, such as Chignolin or BBA, the minimum RMSD macrostate aligns with the native structure, exhibiting an average RMSD of less than 3.0 \AA. Moreover, their equilibrium probabilities are 24.5 $\pm$ 0.1  \% for Chignolin and 17.3 $\pm$ 0.1 \% for BBA (Table \ref{tab:minMacro}). This evidence suggests that our NNP can effectively generalize when trained on an ample dataset. For the other proteins, the minimum RMSD of the macrostates is consistently lower than 5 \AA. As a result, although the macrostates do not precisely match the native structure, near-native conformations are sampled during the simulations with notable probability. 

Simulations for Chignolin, BBA, and Homeodomain proteins successfully sampled folding events, even when starting from completely unfolded conformations. The folding events observed in our simulations, illustrated in Figure \ref{fig:folding_general}, provide compelling evidence that our G-NNP generalizes beyond the training set, accurately folding proteins not included in the training set, albeit with limitations in accurately reproducing the thermodynamic landscapes. As illustrated in Figure S3 of the supplementary information, the free-energy landscapes produced by our models do not reproduce those generated by all-atom MD.

In addition to the aforementioned observations, Figure (\ref{fig:conformations}) illustrates that the G-NNP-selected structures exhibit greater variability than those sampled with the FF-NNP. Nevertheless, the G-NNP more accurately recovers secondary structure elements and the overall shape for proteins ($\lambda$-repressor, a3d, and Protein B) where the FF-NNP fell short. 

\subsection{The Neural Network Potential is comparable to other methods in de-novo structure prediction}
While AlphaFold2\cite{Jumper2021} has indeed revolutionized the domain with its ability to predict native state models, it largely depends on multiple sequence alignments (MSAs) during the initial stages of native state prediction. In contrast, our approach does not require the utilization of such information, positioning it as potentially more versatile, particularly in contexts where obtaining MSAs is a challenge. Additionally, our model holds the potential of not only predict the native structure but also the whole folding process.

In this section, we evaluate the G-NNP performance in de-novo structure prediction, where experimental structures are not available. With this, we evaluate the capabilities of our learned CG forcefield to not only run dynamics but also fold proteins to their native conformation.  For this purpose, we predict structures by selecting the most probable macrostate of the simulations used in the previous section.

To benchmark our model (G-NNP) against other methods, we calculated the average root mean square deviation (RMSD) of the most probable macrostate derived from the simulations. The results are presented in Table \ref{tab:comparison}. We compared our findings with two web servers employing coarse-grained methods for protein folding, UNRES \cite{unres} and CABS-fold \cite{cabs-fold}, as well as the only other method utilizing differentiable molecular simulations to learn coarse-grained parameters, DMS \cite{greener}. For the CABS-fold method, as the server was not operational during our analysis, the results for Chignolin, Trp-cage, BBA, and Villin were obtained from the paper by Greener et al. \cite{greener}, for DMS we run the predictions using the same settings they use in their paper and for the UNRES method, models were generated on their web server  using the parameters provided in the MREMD structure prediction example from their tutorial. 

\begin{table}[tb]
    \centering
    \caption{Comparison of C$\alpha$ RMSDs (\r{A}) 12 fast folding proteins predicted structures with our model and different Coarse-graining models. For our method (G-NNP) we show the mean RMSDs of the most probable macrostate.}
    \resizebox{\columnwidth}{!} {
    \begin{tabular}{ccccc}
        Protein & DMS & UNRES & CABS-fold  & G-NNP (ours)  \\
    \hline
    Chignolin & \textbf{2.7} & 4.8 & 4.8 & 5.28  \\ 
    \hline
    Trp-cage  & 5.6 & \textbf{2.7} & 3.5 & 5.47   \\
    \hline
    BBA  & 3.6 & 7.2 & 7.9 & 7.54 \\
    \hline 
    Villin  & 7.4 & \textbf{6.4} & 11.5 & 7.21 \\
    \hline
    BBL  & 8.68 & 11.6 & -- & \textbf{6.26}  \\
    \hline 
    WW-domain  & 9.07 & 8.7 & -- & \textbf{8.46} \\
    \hline 
    Protein B & 8.88 & 7.1 & -- &  \textbf{6.56} \\
    \hline 
    Protein G  & 11.59 & 11.7 & -- & \textbf{9.76}  \\
    \hline 
    NTL9  & 9.58 & 8.3 & -- & \textbf{7.34}  \\
    \hline
    Homeodomain  & 6.82 & \textbf{5.3} & -- & 6.17  \\
    \hline 
    a3d  & 11.81 & 8.9 & -- &  \textbf{3.73} \\
    \hline 
    $\lambda$-repressor  & 12.48 & \textbf{9.2} & -- &  11.2 \\
    \hline 
    \end{tabular}
    }
    \label{tab:comparison}
\end{table}

Our general model has produced comparable results to other models that use coarse-graining simulations for predicting folded protein conformations. However, CABS-fold and UNRES employ replica exchange algorithms to enhance sampling.  Additionally, the DMS method used an initial guess of the secondary structure as a starting point for the simulations, which may impact the comparability of the results. Nonetheless, we want to emphasize that our neural network model, which was trained from scratch on experimental structures, can achieve results similar to those of more sophisticated, pre-existing, manually crafted methods, or DMS, which are more memory and time-intensive. 

Another aspect of our method, and the ones we have used as a benchmarks (UNRES, CABS-fold and DMS) is their capacity to illustrate not only the end conformation, as current protein structure prediction methods \cite{Jumper2021, Lin2023, Baek2021}, but also the pathway the protein traverses towards it. This aspect could provide a more comprehensive understanding of protein dynamics, and in combination with additional reference data, it could eventually predict both structure and folding pathways.

It is worth noting that our current model does not fully encapsulate the comprehensive reproduction of the entire conformational landscape at this stage. Despite this, we envision our method as a significant stepping stone paving the way for future advancements in the field. Looking ahead, we perceive the potential of this approach to be used as a pre-training stage that can be trained on large amounts of proteins to capture relevant information. Subsequently, it could be combined with active learning strategies to learn the exact forces in sampled conformations, thus more accurately mirroring the thermodynamics, or used as a foundation model that can be fine-tuned for specific downstream tasks.

\section{Conclusions}

In this study, we have effectively extended the application of the differentiable trajectory reweighting algorithm for the parameterization of neural network-based protein force fields. We developed a fast-folders neural network potential (NNP) using 12 proteins, highlighting its ability to fold  and stabilize the native conformations of proteins within the training data distribution. Furthermore, we constructed a general NNP and showed that the learned potential can generalize outside of the training distribution and predict the folded macrostates of proteins with accuracy similar to existing classical coarse-grained methods. Remarkably, the general NNP, while only trained to maintain the native structure, demonstrated the capability to fold some proteins, which their sequence was not present in the training, set starting from entirely unfolded conformations. 

We demonstrated that neural network potentials (NNPs) can be trained in a top-down manner, removing the need for expensive reference calculations or memory-intensive end-to-end differentiable simulations when addressing the protein folding problem. While our current results do not encompass the entire protein folding process, including kinetics and thermodynamics, we are optimistic that future enhancements to our approach, in conjunction with bottom-up methodologies, will enable NNPs to achieve superior accuracy and faster inference times compared to current techniques. Future research may involve integrating our method with labeled data from extensive simulations to create a model capable of accurately predicting protein folding behavior through coarse-grained molecular dynamics simulations. \\

\textbf{Data and code availability} 
\newline
Code, models, prior parameters and all the data are freely available in: \\
\href{https://github.com/compsciencelab/torchmd-exp}{github.com/compsciencelab/torchmd-exp} \\

\textbf{Associated Content}
\newline
Supporting Information available. The SI provides detailed mathematical formulations for the prior energy terms used in the model, as well as the architecture and equations for the Graph Neural Networks. It also includes training curves for different potentials (Figures S1-S2), a comparative analysis of the free energy landscape across initial TICA dimensions (Figure S3), and a table listing the sequences of fast-folding proteins studied (Table S1). Additional tables outline the hyperparameters used for neural network training (Table S2) and a comparison of MD simulation speeds for different proteins using different NNPs (Table S3).

\bibliography{references}

\end{document}


\maketitle
\section*{Prior energy terms}

The pairwise bonded term was:
\begin{equation}
    U_{\lambda}^{harmonic}(r) = k(r-r_0)^2 + V_0
\end{equation}
where $r$ is the distance between beads of a given bond, $r_0$ is the equilibrium distance and $k$ is the spring constant. \\
The nonbonded repulsive term was:
\begin{equation}
    U_{\lambda}^{repulsive}(r) = 4\epsilon r^{-6} + V_0
\end{equation}
where $\epsilon$ is a constant fitted to the data and r is the distance between two beads. Finally, the dihedral prior was:
\begin{equation}
    U_{\lambda}^{dihedral}(\phi) = \sum_{n=1,2}k_n(1 + cos(n\phi - \gamma_n))
\end{equation}
where $\phi$ is the dihedral angle between four consecutive beads, $k_n$ is the amplitude and $\gamma_n$ is the phase offset of the harmonic component of periodicity $n$. The parameters for the priors were fitted to data of all-atom simulations of the fast-folding proteins. For the dihedral prior we used specific dihedrals for each combination of four beads, and for the general NNP we considered all the four beads combinations equally since the straining dataset was bigger. In both cases, the values of $k_n$ were divided by five to achieve a soft prior that does not disturb the simulation too much.

\section*{Graph Neural Networks}

The Graph Neural Network used as an NNP is the same as the one  in the publicly available package TorchMD-NET\cite{torchmdnet}. The network is inspired by SchNet\cite{Schutt2018} and PhysNet\cite{Unke2019}, and optimized for coarse-graining. The network takes as input the Cartesian coordinates of the coarse-grained beads, alongside a predetermined type for each bead. Each bead represents a node of the graph and is given an embedding feature vector by applying a learnable linear mapping. The series of network operations can be written as:

\begin{align*}
    \xi^0 &= W^Ez \\
    \xi^1 &= \xi^0 + W^0\sigma(Aggr(W^C * \xi^0)) \\
    \xi^2 &= \xi^1 + W^1\sigma(Aggr(W^C * \xi^1)) \\
    &\phantom{=} \vdots \\
    \xi^N &= \xi^{N-1} + W^{N-1}\sigma(Aggr(W^C * \xi^{N-1})) \\
    U &= H_{out}(\xi^N) \\
    F &= -\operatorname{grad}(U,x)
\end{align*}

for N interaction layers. Where $W^C$ are continuous filters, generated by expanding the pairwise distance between beads into a set of radial basis functions. $Aggr$ is an aggregation function that reduces the convolution output, in our case we choose the sum as the aggregation method. Finally, the graph level feature $U$ is computed which, in our case, corresponds to the potential energy of the protein and that can be used to compute the forces acting on each bead with an autograd call with respect to the Cartesian coordinates. \\

\section*{Training curves}

\begin{figure}
\centering
\begin{minipage}{.5\textwidth}
  \centering
    \captionsetup{width=.8\linewidth}
  \includegraphics[width=\linewidth]{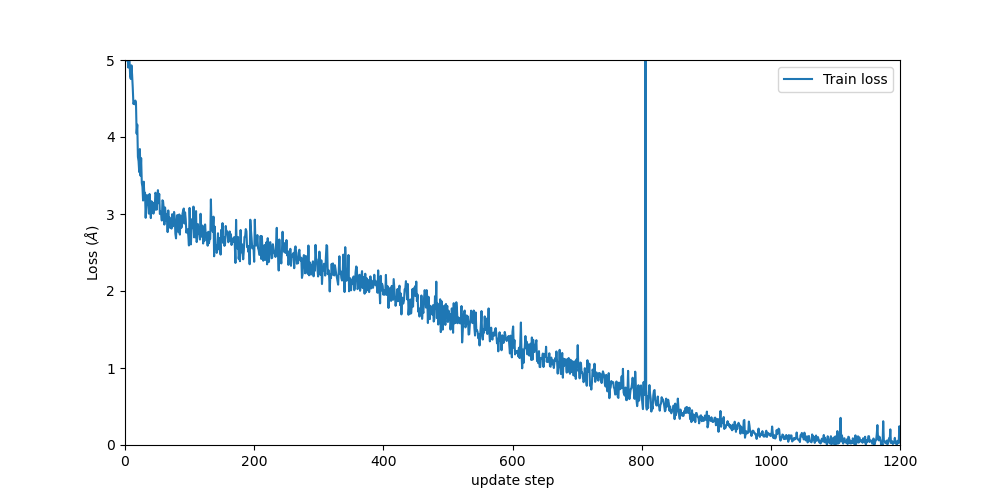}
  \captionof{figure}{Train curve for transferable fast-folding potential.}
  \label{fig:test1}
\end{minipage}%
\begin{minipage}{.5\textwidth}
  \centering
  \captionsetup{width=.8\linewidth}
  \includegraphics[width=\linewidth]{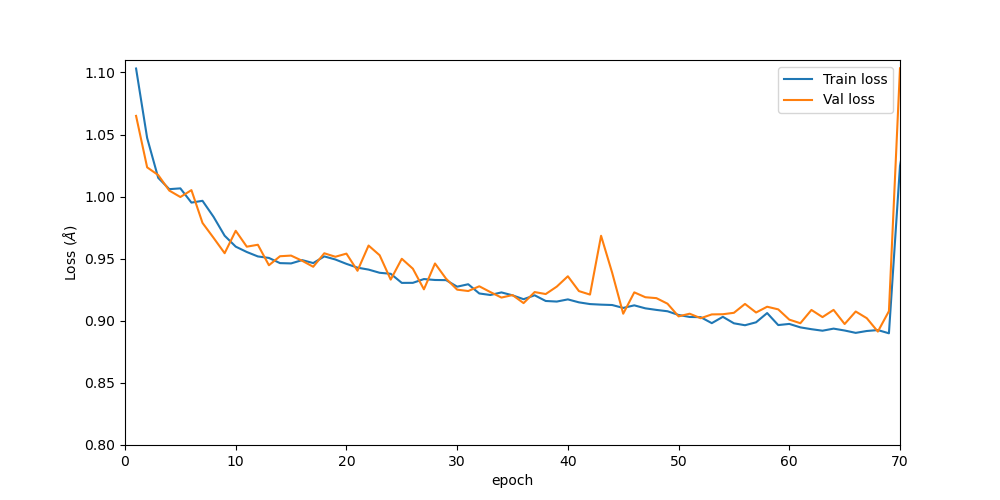}
  \captionof{figure}{Train curve for the general NNP potential.}
  \label{fig:test2}
\end{minipage}
\end{figure}

\section*{Time-lagged independent component analysis}
\begin{figure}
      \centering
      \captionsetup{width=1\linewidth}
      \includegraphics[width=\linewidth]{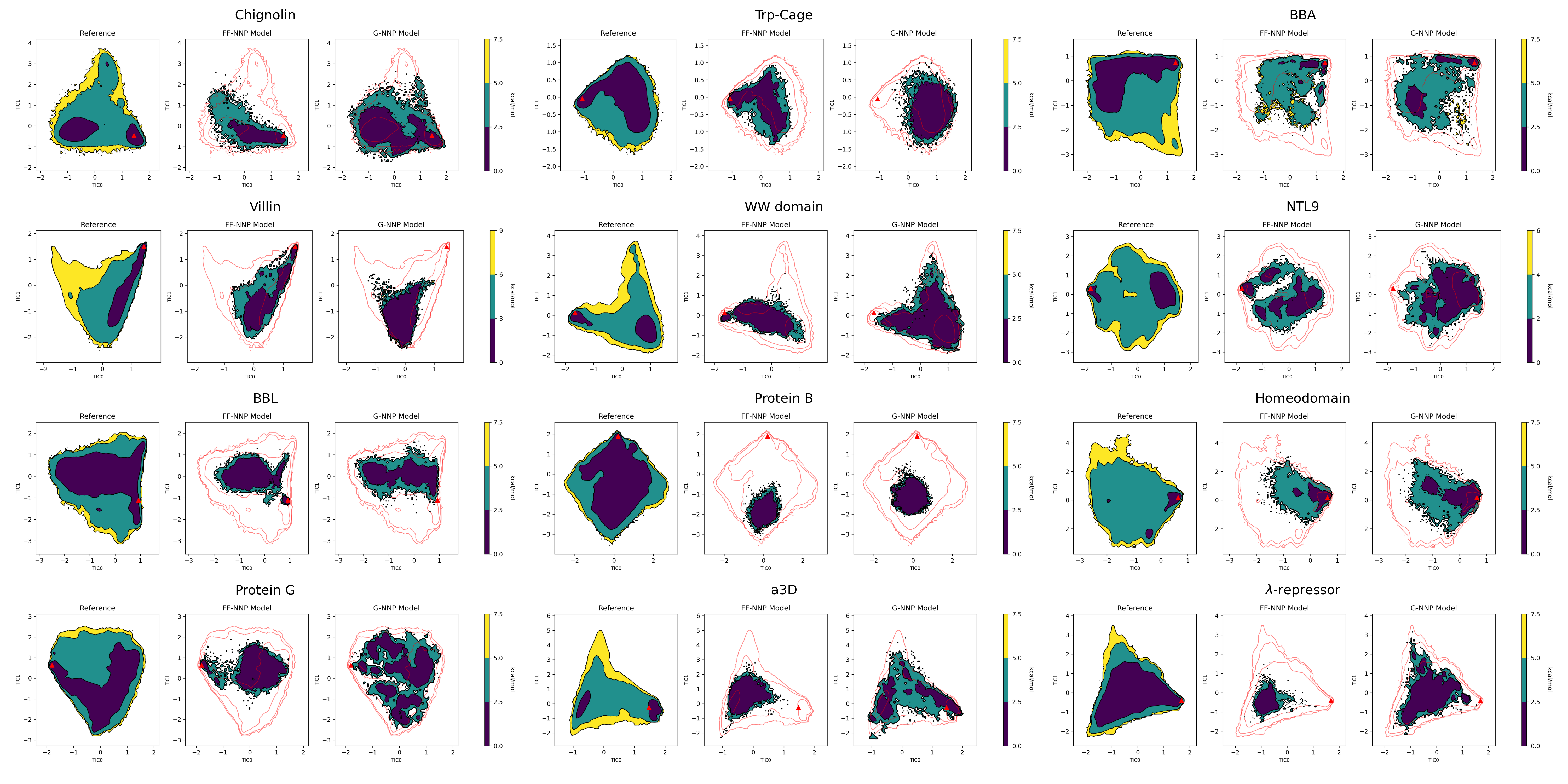}
      \caption{Comparative analysis of the free energy landscape derived from reference all-atom molecular dynamics (MD) (left), FF-NNP model (center), and G-NNP model (right) during coarse-grained simulations across the initial two TICA dimensions for each protein under study. The computation of the free energy landscape for each simulation group involved the segmentation of the first two TICA dimensions into a grid of 80x80 cells, followed by the calculation of the average weights of the equilibrium probability within each cell, as delineated by the Markov state model. The delineated red line marks the energy level above the free energy minimum indicated by the all-atom equilibrium density, characterized by values of 9kcal/mol for Villin and a3d, 6 kcal/mol for NTL9, and 7.5 kcal/mol for the remaining proteins.}
     \label{fig:TICA}
\end{figure}

\begin{table}[H]
    \caption{Fast-folding proteins length and sequence. }
    \resizebox{\columnwidth}{!} {
    \begin{tabular}{l c l c c }
    \hline
    Protein & Length & Sequence   \\
    \hline
    Chignolin & 10 & YYDPETGTWY &  \\ 
    \hline
    Trp-cage  & 20 & DAYAQWLKDGGPSSGRPPPS \\
    \hline
    BBA  & 28 & EQYTAKYKGRTFRNEKELRDFIEKFKGR \\
    \hline 
    WW-domain & 34 & KLPPGWEKRMSRSSGRVYYFNHITNASQWERPSG \\
    \hline
    Villin  & 35 & LSDEDFKAVFGMTRSAFANLPLWXQQHLXKEKGLF \\
    \hline
    NTL9 & 39 &  MKVIFLKDVKGMGKKGEIKNVADGYANNFLFKQGLAIEA\\
    \hline
    BBL & 47 & GSQNNDALSPAIRRLLAEWNLDASAIKGTGVGGRLTREDVEKHLAKA \\
    \hline
    Protein B & 47 &  LKNAIEDAIAELKKAGITSDFYFNAINKAKTVEEVNALVNEILKAHA \\
    \hline
    Homeodomain & 54 & RPRTAFSSEQLARLKREFNENRYLTERRRQQLSSELGLNEAQIKIWFQNKRAKI\\
    \hline
    Protein G & 56 & DTYKLVIVLNGTTFTYTTEAVDAATAEKVFKQYANDAGVDGEWTYDAATKTFTVTE \\
    \hline
    a3D & 73 &  MGSWAEFKQRLAAIKTRLQALGGSEAELAAFEKEIAAFESELQAYKGKGNPEVEALRKEAAAIRDELQAYRHN \\
    \hline
    $\lambda$-repressor & 80 & PLTQEQLEDARRLKAIYEKKKNELGLSQESVADKMGMGQSGVGALFNGINALNAYNAALLAKILKVSVEEFSPSIAREIY\\
    \end{tabular}
    }
\end{table}

\section*{Training hyperparameters}
Different architecture hyperparameters were used for each dataset (Table \ref{tab:hparams}). For the smaller dataset, we used a network with a small cutoff and less number of layers in order to have fewer parameters, which helped to avoid overfitting. 
4 NVIDIA GeForce RTX 2080 machines were used for training the general NNP and  1 NVIDIA GeForce RTX 2080 for the fast-folders dataset.

\begin{table}[H]
    \caption{Hyperparameters choices for the neural networks trained on the fast folders and the monomers datasets. The neural network was implemented with the TorchMD-Net package\cite{torchmdnet}, and the rest of the parameters were left to the default values.}
    \resizebox{\columnwidth}{!} {
    \begin{tabular}{l c l c c }
    \hline
    Hyperparameter & Value for the Fast-Folders NNP  & Value for the General NNP   \\
    \hline
    Number of interaction layers & 1 & 4 &  \\ 
    \hline
    Activation function  & tanh & tanh \\
    \hline
    Radial base function (RBF) type  & expnorm & expnorm \\
    \hline 
    Number of RBF & 18 & 18 \\
    \hline
    Upper cutoff for RBF  & 9.0 & 12.0 \\
    \hline
    Lower cutoff for RBF & 3.0 &  3.0 \\
    \hline
    Trainable RBF & True & True \\
    \hline
    Model type & graph-network &  graph-network \\
    \hline
    Embedding dimension & 256 & 256 \\
    \hline
    Learning rate & 1.0e-4 & 5.0e-4 \\
    \hline
    Neighbor embedding & False &  False \\
    \hline
    Batch size & 12 & 32\\
    \hline
    Total parameters & 326437 & 1129765 \\
    \end{tabular}
    }
    \label{tab:hparams}
\end{table}

\section*{Simulation speed}

Simulations were performed using single NVIDIA GeForce GTX 1080 for the F-NNP and NVIDIA GeForce RTX 4090 for the G-NNP, in Table \ref{tab:speed} we summarize the speed of the two models for each simulated system.

\begin{table}[H]
    \caption{Comparison of MD simulation speed (ns/day) of the FF-NNP and G-NNP. The results obtained with NVIDIA GTX 1080 for the FF-NNP and with NVIDIA GeForce RTX 4090 for the G-NNP.}
    \begin{tabular}{l c l c c }
    \hline
     Protein   & Fast-Folders NNP (ns/day)  & General NNP (ns/day)   \\
    \hline
    Chignolin & 21.48 & 50.56 \\ 
    \hline
    Trp-cage  & 20.69 & 52.66 \\
    \hline
    BBA  & 20.56 & 50.24 \\
    \hline 
    WW-domain & 21.11 & 49.98 \\
    \hline
    Villin  & 21.24 & 51.89 \\
    \hline
    NTL9 & 21.40 & 50.20 \\
    \hline
    BBL & 20.62 & 48.89 \\
    \hline
    Protein B & 21.00 & 47.66  \\
    \hline
    Homeodomain & 11.52 & 49.23\\
    \hline
    Protein G & 9.97 & 47.60 \\
    \hline
    a3D & 9.32 & 45.70 \\
    \hline
    $\lambda$-repressor & 8.70 & 44.08 \\
    \end{tabular}
    \label{tab:speed}
\end{table}

\bibliography{references}